\documentclass[12pt]{elsart}
\usepackage[dvips]{graphicx}
\usepackage{amsmath,amssymb}
\begin{document}

%%%%%%%%%%%%%%%%%%%%%%%%%%%%%%%%%%%%%%%%%%%%%%%%%%%%%%%%%%%%%%%%%%%%%%
\begin{frontmatter}

\title{Linear morphological stability analysis 
of the solid-liquid interface
in rapid solidification of a binary system}

\author{P.~K.~Galenko*\thanksref{Correspond}}
\and \author{D.~A.~Danilov**}
\address{*German Aerospace Center, 
Institute of Space Simulation, 51170 Cologne, Germany \\
  **University of Applied Sciences,
   76133 Karlsruhe, Germany}

\thanks[Correspond]{\mbox{Fax: ++49-2203-601-2255;}
e-mail:~Peter.Galenko@dlr.de}

\begin{abstract}
  The interface stability against small perturbations of the planar
  solid-liquid interface is considered analytically in linear
  approximation. Following the analytical procedure of Trivedi 
  and Kurz (Trivedi R, Kurz W. Acta Metall 1986;34:1663), 
  which is advancing the original treatment of morphological 
  stability by Mullins and Sekerka (Mullins WW, Sekerka 
  RF. J Appl  Phys 1964;35:444) to the case of rapid 
  solidification, we extend the model by introducing the 
  local nonequilibrium in the solute diffusion field around 
  the interface. A solution to the heat- and mass-transport 
  problem around the perturbed interface is given 
  in the presence of
  the local nonequilibrium solute diffusion. Using the developing
  local nonequilibrium model of solidification, the self-consistent 
  analysis of linear 
  morphological stability is presented with the attribution
  to the marginal (neutral) and absolute morphological stability of
  a rapidly moving interface. Special consideration of the
  interface stability for the cases of solidification in 
  negative and positive thermal gradients is given. A quantitative
  comparison of the model predictions for the absolute morphological
  stability is presented with regard to experimental results of Hoglund 
  and Aziz (Hoglund DE, Aziz MJ. Mat Res Soc Symp Proc 1992;205:325) 
  on critical solute concentration for the interface breakdown during
  rapid solidification of Si--Sn alloys.  \linebreak 

\end{abstract}
\end{frontmatter}
%%%

\newpage
%%% INTRODUCTION %%%%%%%%%%%%%%%%%%%%%%%%%%%%%%%%%%%%%%%%%%%%%%%%
\section{Introduction} \label{sec:intro}
In the solidification of liquids, an initial solid-liquid interface 
is subject to unstable growth which leads to various crystal patterns 
of cellular, dendritic, banded, fractal, etc. morphologies 
\cite{Flem,chernov,vicsek,Kurz-Fisher,PhysDendr}.  
A scheme of changing the crystal morphology with the interface velocity 
can be considered in the example of one-phase solidification, i.e. when 
the liquid transfers into solid without precipitation of additional phases. 
Figure \ref{fig:morph_spectr-1} shows schematically the steady-state 
growth morphologies which form in a liquid as a result of morphological 
instability at given interface velocity, $V$, in single-phase 
solidification. 
With a small velocity, an initially smooth interface 
remains planar up to a velocity equal to the critical velocity, $V_C$, defined 
by the constitutional undercooling \cite{cons}. 
Behind $V_C$, the smooth interface becomes unstable and the interface 
exhibits a steady cellular morphology. By further increasing the velocity, 
a surface of cells may become unstable with the developing 
of dendritic patterns. At high interface velocity, dendritic patterns degenerate 
with the appearing of rapidly moving cells. A demarcation line at $V=V_A$ 
divides the regions between the interface instability, $V<V_A$, 
and the absolute stability, $V>V_A$, where the planar interface is 
morphologically stable against small perturbations of its form. 
This demarcation line is usually known as the critical 
velocity, $V_A$, for absolute stability of the planar interface. 
The sequence of growth morphologies, Fig.~\ref{fig:morph_spectr-1}, 
is well-known from experiments on directional solidification 
and solidification in the undercooled state \cite{sci-tech}. 
It has been demonstrated 
in computational modeling \cite{gk} of crystal growth as well. 

The theory of morphological stability was developed first 
by Mullins and Sekerka, who 
considered the stability of a spherical particle grown into a supersaturated 
solution \cite{ms1}, 
and stability of the planar front during directional solidification 
of a binary liquid \cite{Mullins-Sekerka-1964}. 
In investigating the evolution of small harmonic 
perturbations of the planar interface \cite{Mullins-Sekerka-1964}, they 
provided a rigorous basis of linear morphological (in)stability at low 
solidification velocity. Particularly, Mullins and Sekerka  
introduced a concept of marginal stability for the wavelength of 
perturbation, which gives neutral stability of the plane when  
the amplitude of perturbation does not change in time. Behind the critical velocity 
of the marginal stability, the increasing of the amplitude of 
perturbation in time may lead to cellular or dendritic interfaces. 
The theory \cite{Mullins-Sekerka-1964} gave rise to 
a great number of investigations of morphological transformations due to linear 
instability of interfaces and nonlinear behaviour of unstable interfaces. 
These are presented in overview by  
Coriell and McFadden \cite{cor-mac} and monograph by Davis 
\cite{davis}. 

In its classic form \cite{Mullins-Sekerka-1964}, the theory assumes 
a local equilibrium at the interface, which is an excellent approximation 
for many metallic systems solidifying at small interface velocities. 
At large driving force for the interface advancing, and 
with the increasing of its velocity, the analysis 
of Mullins and Sekerka can be modified. 
Trivedi and Kurz \cite{Trivedi-Kurz} extended the analysis 
of Mullins and Sekerka \cite{Mullins-Sekerka-1964} for the case 
of rapid solidification, and introduced the stability functions 
dependent on the interface velocity. By taking into account the velocity 
dependent coefficient of solute redistribution (partitioning function), 
they developed an analytical model 
\cite{Trivedi-Kurz-2} of microstructure formation under directional 
solidification, over the range from low interface velocity up to velocity, $V_A$, 
of absolute morphological stability. 

In addition to the previous analysis of pattern formation and 
morphological stability of the interface in   
which the treatment is extended to rapid solidification \cite{Trivedi-Kurz} 
and nonequilibrium effects at the interface \cite{Trivedi-Kurz-2}, 
the local nonequilibrium in bulk phases may play an essential 
role in rapid solidification. Particularly, a deviation from local 
equilibrium in solute diffusion may act on the rapid advancing 
of the solid-liquid interface because the interface velocity, $V$, 
can be of the order or even greater than the solute diffusion speed, 
$V_D$, in bulk phases. For instance, the diffusion 
speed can be of the order of \cite{G1}: $V_D\sim 0.1-10$~m/s. 
In modern experiments on solidification of undercooled droplets 
the interface velocity approaches \cite{herl}: 
$V=10-100$~m/s. Therefore, the undercooling of liquids  
is sufficient for detecting solidification with the interface velocity 
comparable to the diffusion speed. 

Considering the process of solute diffusion, Fick's first 
law is obtained on the basis of classic irreversible 
thermodynamics of Onsager and Prigogine 
which assumes propagation of 
concentration disturbances with an infinite speed and 
with local equilibrium in the bulk phases \cite{pr1}. Local 
equilibrium is characterized by the statistical 
distribution function, given by the first order term 
of its expansion \cite{prig}. However, for a high-velocity 
solidification front, the time for crystallizing of 
a local volume is comparable to the time for 
relaxation of the diffusion flux to its steady-state value \cite{G1}. 
In this case, 
local equilibrium is absent in the bulk phases and the 
solute flux cannot be described by the classical Fick's 
first law. Including the evolution equation for the diffusion flux, 
the analysis of Galenko and Sobolev \cite{GS} shows 
that the deviations from local equilibrium in phases and at the interface 
drastically affect both the solute diffusion and the interface kinetics. 
Therefore, in this article we 
consider the linear stability analysis for a rapidly moving interface 
under local nonequilibrium solute diffusion. 

The linear morphological analysis of the interface stability 
has been developed and used to obtain a stable tip of 
dendrite growing under local nonequilibrium 
solute diffusion in rapid solidification \cite{GD1}. A 
marginal stability criterion was used and concluded that 
at $V=V_D$ the complete transition to 
diffusionless solidification may proceed sharply, with the appearing of 
the break point in the kinetic curves ``dendrite velocity - undercooling" 
and ``dendrite tip radius - undercooling". In addition to this, Lee et al. 
\cite{lee} performed the linear stability analysis in rapid directional 
solidification using the model of Galenko and Sobolev \cite{GS}.  
As they showed \cite{lee}, the effect of the local nonequilibrium in solute 
diffusion postpones the onset of the cellular instability in better 
agreement with experimental data, in comparison with the model predictions 
in which the local nonequilibrium only at the interface is considered. 
However, the analysis of Lee et al. \cite{lee} was not self-consistent 
(in the analysis, the authors used an expression for the slope of 
kinetic liquidus obtained from the local equilibrium thermodynamics). 
As it has been noted in Ref.~\cite{GD1}, the predictions of rapid 
solidification of alloys can be satisfactory compared with 
experimental data only on the basis of self-consistent model 
(i.e., when all model functions are taking into account the 
deviation from local equilibrium in the solute diffusion field). 
Consequently, {\it the first purpose} of the present article is 
to analyze the morphological stability of the planar interface 
on the basis of the self-consistent model of local nonequilibrium 
solidification. 

In the present analysis, 
special consideration is paid to the stability of the interface 
around the demarcation line, $V=V_A$, 
below which the interface instability occurs 
and behind of which the absolute morphological stability  
proceeds, Fig.~\ref{fig:morph_spectr-1}. 
This transition is known from experiment as a transition 
from cellular patterns to the segregation-free patterns \cite{cantor,Bo-Co86}. 
The crystal microstructure after the transition is shown in Fig.~\ref{fig:spinning}. 
As it has been stated \cite{gal-lad},   
the transition from the macroscopically smooth 
solid-liquid interface to the cellular-dendritic microstructures occurs 
with the decreasing of the interface velocity, $V$, below critical 
velocity, $V_A$, for absolute stability of the planar front. Around the 
velocity $V=V_A$, the kinetics of crystal growth begin to 
disagree with the predictions of the model in which the local 
nonequilibrium only at the interface is considered (see the results of 
analysis in Ref.~\cite{GD1}). Consequently, 
{\it the second purpose} of this article is a quantitative 
evaluation of the discrepancy between the present local 
nonequilibrium model and the model in which the local 
nonequilibrium only at the interface is taken. It is also in 
comparison with experimental data on morphological 
stability of the interface in solidifying alloy. 
 
The article is organized as follows. In Sec. 2 we give a 
formal description of departure from local equilibrium 
due to solute diffusion and give a set of governing 
equations to analyze the morphological stability of 
the interface. In Sec. 3 we analyze the influence of 
perturbations in fields and at the interface on the 
linear stability of the planar interface. In Sec. 4, 
an obtained criterion of marginal (neutral) stability 
allows us to analyze the morphological stability 
of the interface for the case of solidification an 
undercooled melt (with negative temperature 
gradient) and for the directional solidification 
(with the positive temperature gradient). The 
absolute stability of the planar interface is analyzed in Sec. 5.  
A discussion about expressions for solute 
trapping and kinetic liquidus which define 
a final form of the function for the absolute 
stability of the planar interface is given in Sec. 6. 
Also, in this section, we compare the 
derived function for the absolute interface 
stability with the available experimental results 
obtained in Si-Sn alloy solidification. Finally, 
in Sec. 7 we present a summary of our conclusions. 

%%% STATEMENT OF THE PROBLEM %%%%%%%%%%%%%%%%%%%%%%%%%%%%%%%%
\section{Statement of the problem}
We shall consider a dilute binary alloy that undergoes
nonisothermal solidification in infinite space. Let us take into
account the heat diffusion in phases, 
solute diffusion in the liquid, and one can neglect the solute 
diffusion in solid. The main physical
assumption of the present problem is an absence of local
equilibrium both at the solid-liquid interface and in the solute
diffusion field around the interface. 
In this case, the degree of local
nonequilibrium is estimated by the relation of the interface velocity,
$V$, and the diffusion speed, $V_D$, which is a 
parameter of the process of diffusion and can have
different values at the interface and in bulk phases. The
speed $V_D$ is a maximum speed of propagation of the diffusion profile
in the system and defined as $V_D=(D/\tau_D)^{1/2}$, where
$D$ is the diffusion coefficient, and $\tau_D$ is the time of relaxation
of diffusion flux to its steady-state value. Therefore, 
we develop the rapid solidification model which is 
taking into account the finiteness of the diffusion 
speed in the system.

\subsection{Departure from local equilibrium}
If local thermodynamic equilibrium in the bulk is 
not reached, the connection between 
the vectors of diffusion fluxes, $\bold q_i$ and $\bold J$, 
and the driving forces, $\nabla T_i$ and $\nabla C$, 
for the heat and solute diffusion, respectively, 
have the following integral form: \\
- \textit{relaxation of the heat flux}  
\begin{equation}
\bold q_i(\bold r,t)=-\int_{-\infty}^{t} 
D_q^i(t-t^*)\nabla T_i(t^*,\bold r)dt^*,
\label{991A}
\end{equation}
- \textit{relaxation of the solute diffusion flux}  
\begin{equation}
\bold J(\bold r,t)=-\int_{-\infty}^{t} 
D_j(t-t^*) \nabla C(t^*,\bold r)dt^*,
\label{991}
\end{equation}
where index $i=L$ or $i=S$ is related to the liquid or solid phases, 
respectively. $T_i$ are the temperatures in the phases, 
$C$ is the solute concentration in the liquid, 
$t$ is the time, $\bold r$ is the radius-vector of a point in the
system, and $D_R(t-t^*)$ are the relaxation functions of the 
fluxes ($R=q$ or $R=j$). 
Equations (\ref{991A}) and (\ref{991})
imply the fact that when the
interface moves with a high velocity, local
equilibrium in the fields does not occur and the diffusion fluxes at
a point in the system no longer depend on the instantaneous gradients 
of the temperature and chemical composition, but are also determined by the local
prehistory of the solidification process. 

Equations (\ref{991A}) and (\ref{991}) represent general expressions 
for evolution prehistory of the diffusion processes. For the case of 
heat diffusion, when the heat propagates with much higher speed 
in comparison with the interface velocity, 
the influence of local nonequilibrium in the temperature field on the 
kinetics of the interface advancing is negligible  
(see the analysis of heat transfer in rapid solidification 
in Ref.~\cite{GD_pla_2000}). Therefore  
we specially define the relaxation functions $D_R$ in  
Eqs.~(\ref{991A}) and (\ref{991}) for the important class 
of dissipative hyperbolic models in which 
they take the following forms 
\begin{equation}
D_q^i(t-t^*)=D_q^i(0)\delta(t-t^*), 
\label{26cab}
\end{equation}
\begin{equation}
D_j(t-t^*)=D_j(0)\exp \Big(-\frac{t-t^*}{\tau_D}\Big), 
\label{26bac}
\end{equation}
where $D_q^i(0)=K_i$ are the thermal conductivity in the liquid ($i=L$) 
and solid ($i=S$), $\delta$ is the Dirac delta-function, and 
$D_j(0)=D/\tau_D$ is a value of relaxation function for solute diffusion 
at a moment $t=t^*$.

Eq.~(\ref{26cab}) describes an instant relaxation which occurs at a moment $t=t^*$. 
Therefore, one can expect a description of local equilibrium heat 
transport by the function (\ref{26cab}) in combination with Eq.~(\ref{991A}). 
In contrast to this, 
the relaxation function (\ref{26bac}) with the flux prehistory 
(\ref{991}) leads to local nonequilibrium solute diffusion. 
In usual circumstances, the relaxation time, $\tau_D$, for diffusion flux is very 
small and the relaxation effects are negligible. It is usual to 
speak in this case of "viscous diffusion" which can be accurately 
described by the Fickian local equilibrium approximation. However, 
equations (\ref{991}) and (\ref{26bac}) describe relevant 
difference of solute diffusion with respect to the classical Fick's law. 
To date, departures from this law are known as the appearance of 
internal effects, couplings of diffusion and viscosity, and longitudinal 
diffusion (see Refs.~\cite{jcg,j1,j2,j} and references therein). In addition 
to these appearances, Eqs.~(\ref{991}) and (\ref{26bac}) can 
also be applied to the process of the phase transition in a strongly 
nonequilibrium medium. Rapid solidification of a 
binary melt is a good example of such a nonequilibrium phase 
transition in which the interface can move with the high velocity 
comparable with the diffusive speed, $V\sim V_D\sim 0.1-10$ (m/s) \cite{G1}. 
In this case, 
the time for crystallizing of local bulk is of the order of the relaxation 
time of the solute diffusion flux \cite{G1,GS,GD1}, characterizing 
its decay forward of its local equilibrium value. Therefore when 
$V\sim V_D$, the relaxation {\em interacts} with the diffusion process 
directly, and it is necessary to take into account the local prehistory 
of the solute diffusion, e.g. in a form described by Eqs.~(\ref{991}) 
and (\ref{26bac}). 

Eq.~(\ref{26bac}) simulates a physically reasonable 
situation in which exponential decay of the diffusion flux occurs 
in the local bulk of the liquid phase. This equation provides 
the lowest order of approximation of the diffusion flux relaxation.
Indeed, substituting Eqs.~(\ref{26cab}) and (\ref{26bac}) 
for Eqs.~(\ref{991A}) and (\ref{991}), respectively, one 
obtains 
\begin{equation} \label{eq:q}
\bold q_i + K_i \nabla T_i = 0, 
\end{equation}
\begin{equation} \label{eq:J}
\tau_D \frac{\partial \bold J}{\partial t} + \bold J + D \nabla C = 0.
\end{equation}
Equation (\ref{eq:q}) is a well-known Fourier law which is true for infinite 
thermal speed in the system, i.e. the heat diffusion flux is instantly 
relaxed to its local equilibrium value and the effects of local 
nonequilibrium in the thermal field are negligible. 
Equation (\ref{eq:J}) can be treated as the simplest generalization of
the classical Fick's first law $\bold J+D\nabla C=0$ that is recovered
when $\tau_D=0$ or in stationary situations in which $\partial \bold J
/\partial t =0$. The evolution equation (\ref{eq:J}) takes into account
the relaxation to local equilibrium of the mass flux and is known as
the Maxwell-Cattaneo equation in the context of heat transport
\cite{j1,j2,j}. 
By taking the relaxation (\ref{991}) with the exponential law (\ref{26bac}), 
it follows from Eq.~(\ref{eq:J}) that the flux $\bold J$ at a point in the system 
is defined by the evolution of the concentration gradient 
$\nabla C(t^*,\bold r)$ during the period $t-\tau_D < t^* <t$, 
but not by the gradient $\nabla C(t,\bold r)$ at the moment $t$, 
as in local equilibrium approximation. 
Thus, taking into 
account the exponential decay for the relaxation of the diffusion flux 
(see Eq.~(\ref{26bac})), the simplest evolution equation (\ref{eq:J}) 
is obtained for the interaction of the relaxation process and solute 
diffusion. Instead of Eq.~(\ref{eq:J}), by taking suitable relations for 
the functions of $D(C)$, $D(\partial C/\partial t)$, or $D(\nabla C)$, one 
can describe more complicated situations for non-Fickian diffusion in 
nonequilibrium media. These are described by the evolution equation 
with the higher order time derivatives or couplings of relaxation to 
non-local effects for transient processes \cite{j2}.

The heat and solute diffusion are governed by the balance laws 
\begin{equation} \label{eq:balancet}
\chi_i \frac{\partial T_i}{\partial t} + \nabla \cdot \bold q_i = 0,
\end{equation}
\begin{equation} \label{eq:balance}
\frac{\partial C}{\partial t} + \nabla \cdot \bold J = 0,
\end{equation}
where $\chi_i$ are the heat capacities in phases. 
Substitution of Eqs.~(\ref{eq:q}) and (\ref{eq:J}) into 
Eqs.~(\ref{eq:balancet}) and (\ref{eq:balance}), respectively, gives 
the following system of equations
\begin{equation} \label{(1t)}
\frac{\partial T_i}{\partial t} = a_i\nabla^{2}T_i, 
\end{equation}
\begin{equation} \label{(1)}
\tau_{{D}}\frac{\partial^{2} C}{\partial t^{2}} + 
\frac{\partial C}{\partial t} = D\nabla^{2}C, 
\end{equation}
where $a_i$ are the thermal diffusivities in the liquid ($i=L$) and solid ($i=S$). 
Equation (\ref{(1t)}) is the common partial differential equation of 
parabolic type for the heat transfer which adopts the infinite 
thermal speed. Equation (\ref{(1)}) shows that  Eqs.~(\ref{eq:J}) 
and (\ref{eq:balance}) give rise to the partial differential 
equation of a hyperbolic type for the solute concentration, 
which is the simplest mathematical model combining the
diffusive (dissipative) mode and the propagative (wave) mode of mass
transport under local nonequilibrium conditions. In such a case,
Eq.~(\ref{(1)}) describes the transport process under non-Fickian 
diffusion.

After integration of Eqs.~(\ref{(1t)}) and (\ref{(1)}) over an infinitesimal zone 
that includes the interface, the following boundary conditions for the diffusion 
transport hold 
\begin{equation} \label{(2t)}
-K_L\nabla_{{n}}T_L+K_S\nabla_{{n}}T_S=QV_{{n}},
\end{equation}
\begin{equation} \label{(2)}
-D\nabla_{{n}}C=(C-C_{{S}})V_{{n}}
+\tau_{{D}}\frac{\partial}{\partial t}\big((C-C_{{S}})V_{{n}}\big),
\end{equation}
where $Q$~is the latent heat of solidification, $\nabla_{{n}}T_i$ and 
$\nabla_{{n}}C$ are the normal gradients of temperature and 
solute concentration to the interface, respectively, 
 $V_{n}$ is the normal velocity of the
interface, $C_{S}$ is the solute concentration at the interface in 
the solid phase given by expression
\begin{equation} \label{(3)}
C_{{S}}=kC,
\end{equation}
and $k$ is the coefficient of solute partitioning at the interface. 

\subsection{Governing equations}
Under the assumptions drawn in Section 2.1  we consider 
the governing equations for analysis of the morphological 
stability of the planar interface against small perturbations of its
form. Our analysis is based on the analysis given by 
Trivedi and Kurz \cite{Trivedi-Kurz} which is advancing the treatment of
Mullins and Sekerka \cite{Mullins-Sekerka-1964} to the case of rapid
solidification. 

The analysis of interface stability is given further in the
reference frame moving with the constant velocity $V$. We shall
consider the case when the planar interface given by
equation $z(x)=0$ moves along the $z$-axis of the Cartesian coordinate
system ${(x, z)}$. Then the 2D steady-state fields of the solute
concentration and temperature are obtained 
from Eqs.~(\ref{(1t)})  and (\ref{(1)}) as 
\begin{equation}
 \frac{\partial^2 C}{\partial x^2} + \left( 1- \frac{V^2}{V_D^2} \right)
        \frac{\partial^2 C}{\partial z^2}
   + \frac{V}{D}\frac{\partial C}{\partial z}=0,
  \label{eq:ms-steady-state-C}
\end{equation}
\begin{equation}
\frac{\partial^2 T_L}{\partial x^2} + \frac{\partial^2 T_L}{\partial z^2}
    + \frac{V}{a_L}\frac{\partial T_L}{\partial z}=0,
  \label{eq:ms-steady-state-T_L}
\end{equation}
\begin{equation}
  \frac{\partial^2 T_S}{\partial x^2} + \frac{\partial^2 T_S}{\partial z^2}
  + \frac{V}{a_S}\frac{\partial T_S}{\partial z}=0.
  \label{eq:ms-steady-state-T_S}
\end{equation}

Following the standard procedure of the analysis of 
the morphological stability, let's place harmonic 
perturbation on the planar interface. The perturbation is described by
\begin{equation}
  z \equiv \phi (x,t) = \delta (t) \sin(\omega x),
  \label{eq:ms-perturbation}
\end{equation}
where $\delta$~ is a small amplitude of perturbation (${|\delta|\ll
  1}$), ${\omega = 2\pi /\lambda}$~ is the cyclic frequency with the
wavelength, $\lambda$.  The response functions on the perturbed
interface $\phi(x,t)$, i.e. the temperature, $T_\phi$, and solute
concentration, $C_\phi$, are defined by the following relation
\begin{equation}
  T_\phi = T_m + m C_\phi + \Gamma K,
  \label{eq:ms-Tf}
\end{equation}
where $m$ is the slope of the liquidus line in the kinetic phase diagram 
(i.e., phase diagram for nonequilibrium solidification of a binary system), 
$\Gamma$ is the Gibbs-Thomson coefficient (i.e. the capillary parameter 
defined by the surface energy of the interface), 
and $K$~ is the mean curvature of the perturbed interface.

For the sake of simplicity of the following analysis, 
the kinetic term $V/\mu$ (in which $\mu$ is the kinetic coefficient of atomic
attachment to the interface) is omitted in Eq.~(\ref{eq:ms-Tf}). 
This simplification has no influence
on the main results of the present analysis due to the fact that 
a constant kinetic coefficient does not affect the marginal condition 
of the front stability \cite{cor-mac,li}. It has been shown that the 
kinetic effects influence the interface stability in the case where temperature 
dependence of the kinetic coefficient \cite{li}, or when an 
optimum stability conjecture for selection of the 
interface operating state, is used \cite{sekerka}. 

Considering only the terms of the first order of smallness by the
amplitude $\delta$ of perturbation, one can obtain for the curvature
of the perturbed interface the following expression
\begin{equation}
  K = \left( \frac{\partial^2 \phi}{\partial x^2} \right)
  \left(
    1 + \left(
      \frac{\partial \phi }{\partial x}
    \right)^2
  \right)^{-3/2}
  = -\delta \omega^2 \sin(\omega x).
  \label{eq:ms-K}
\end{equation}
Thus, the system (\ref{eq:ms-steady-state-C})-(\ref{eq:ms-K}) is 
the extension of the model for the case of significant deviations 
from local equilibrium in the solute diffusion field during rapid 
solidification. In the case of the local equilibrium solute diffusion, 
i.e. as $V_D \rightarrow \infty$, Eq.~(\ref{eq:ms-steady-state-C}) 
describes the approximation of Fick's diffusion 
which has been used in the existing models (see, e.g. 
Ref.~\cite{Kurz-Fisher}). The system 
(\ref{eq:ms-steady-state-C})-(\ref{eq:ms-K}) has been used in the 
description of local nonequilibrium solidification with planar and 
non-planar solid-liquid interfaces \cite{GS,GD1,GD-99,GD-pla-parab}. 
With the definition of the functions for solute partitioning $k(V)$ and 
the liquidus slope $m(V)$, the system of 
Eqs.~(\ref{eq:ms-steady-state-C})-(\ref{eq:ms-K}) 
can be also applied to the problem of morphological stability 
of the interfaces in rapid solidification. 

%%% ANALYSIS OF MORPHOLOGICAL STABILITY %%%%%%%%%%%%%%%%%%%%%%%%%
\section{Morphological stability}
Within the linear analysis of stability,
a solution of equations (\ref{eq:ms-steady-state-C})--(\ref{eq:ms-steady-state-T_S})
on the perturbed interface (\ref{eq:ms-perturbation}) is described by
\begin{equation}
  C_\phi = C_0 + b\delta (t) \sin (\omega x),
  \label{eq:ms-C_phi}
\end{equation}
\begin{equation}
  T_\phi = T_0 + a\delta (t) \sin (\omega x),
  \label{eq:ms-T_phi}
\end{equation}
where $T_0$ and $C_0$~ are the values for the temperature and solute concentration
for the unperturbed planar interface, respectively, and the parameters $b$ and $a$ define
the corresponding corrections to the small perturbations on it.
Taking into account the small magnitude of perturbations on the planar interface, 
the perturbed steady-state solution for solute concentration can also be
presented in a form proportional to ${\delta(t)\sin(\omega  x)}$.
This leads to the following expression
\begin{equation}
  \label{eq:ms-C(x,z)}
  C(x,z) = \overline{C}(z) + F(z) \delta (t) \sin (\omega x),
\end{equation}
where $\overline{C}(z)$ is the solute distribution for the planar interface.
The function $F(z)$ is defined from Eq.~(\ref{eq:ms-C_phi}). The far-field
condition takes the value for the planar interface, 
i.e. ${F(z)\rightarrow 0}$ with ${z\rightarrow\infty}$.

Substitution of Eq.~\eqref{eq:ms-C(x,z)} into solute diffusion equation
(\ref{eq:ms-steady-state-C}) leads to the following approximations: \\
- in the zero order of magnitude by the amplitude of perturbation $\delta$
\begin{equation}
  \label{eq:ms-delta-0}
  (1-V^2/V_D^2)
  \frac{d^2 \overline C}{d z^2}
  + \frac{V}{D} \frac{d \overline C}{d z}
  = 0,
\end{equation}
- in the first order of magnitude by the amplitude of perturbation $\delta$
\begin{equation}
  \label{eq:ms-delta-1}
  (1-V^2/V_D^2)
  \frac{d^2 F}{d z^2}
  + \frac{V}{D} \frac{d F}{d z}
  - \omega^2 F
  = 0.
\end{equation}

A general solution of Eq.~\eqref{eq:ms-delta-0} has the following form
\begin{equation}
  \label{eq:ms-orevline-C-generic}
  \overline{C}(z)
  = C_1
  + C_2 \exp \left(
    - \frac{V z}{D(1-V^2/V_D^2)}
  \right).
\end{equation}
Solution \eqref{eq:ms-orevline-C-generic} must be limited at ${z \rightarrow \infty}$
and it satisfies the following conditions:
${\left.\overline{C}\right|_{z=0}=C_0}$ and
${\left.d\overline{C}/dz\right|_{z=0}=G_C}$, where 
$G_C$ is the
concentration gradient at the unperturbed interface.
Solution \eqref{eq:ms-orevline-C-generic} satisfies these conditions
in the following form
\begin{equation}
  \label{eq:ms-orevline-C}
  \overline{C}(z) =
  \begin{cases}
    C_0 + \displaystyle\frac{G_C D (1-V^2/V_D^2)}{V}
    \left[
      1 -
      \exp \left(
        - \displaystyle\frac{V z}{D(1-V^2/V_D^2)}
      \right)
    \right], & V<V_D, \\
    C_0, & V \geqslant V_D.
  \end{cases}
\end{equation}

A general solution of Eq.~(\ref{eq:ms-delta-1}) has the following form
\begin{equation}
  \label{eq:ms-F-generic}
  \begin{split}
    F(z) &
    = F_0 \exp
    \left(
      - \frac{V +
        \big(V^2+4D^2(1-V^2/V_D^2)\omega^2 \big)^{1/2}}%
      {2D(1-V^2/V_D^2)} z
    \right) \\
    & + F_1 \exp
    \left(-
      \frac{ V -
        \big( V^2+4D^2(1-V^2/V_D^2)\omega^2 \big)^{1/2}}%
      {2D(1-V^2/V_D^2)} z
    \right).
  \end{split}
\end{equation}
Solution \eqref{eq:ms-F-generic} must be limited at
${z\rightarrow\infty}$ and it takes only the real values for any $\omega$.
In this case, Eq.~\eqref{eq:ms-F-generic} leads to the following particular solution 
\begin{equation}
  \label{eq:ms-F}
  F(z) =
  \begin{cases}
    F_0 \exp \left(
        - \displaystyle\frac{\omega_C z}{\big(1-V^2/V_D^2 \big)^{1/2}}
      \right), & V<V_D, \\
    0, & V \geqslant V_D.
  \end{cases}
\end{equation}
Here $F_0$ is a constant of integration, and the frequency $\omega_C$
is related to the frequency $\omega$ of the perturbation as
\begin{equation}
  \label{eq:ms-omega_C}
  \omega_C =
  \frac{V}{2D(1-V^2/V_D^2)^{1/2}}
  + \left[
    \left(
      \frac{V}{2D(1-V^2/V_D^2)^{1/2}}
    \right)^2
    + \omega^2
  \right]^{1/2}.
\end{equation}
Note that once the interface velocity is equal to or greater than the diffusion speed,
${V\geqslant V_D}$, solution \eqref{eq:ms-F-generic} takes
the zero values for both constants of integration. ${F_0=0}$ is due to
limiting of the solution at the infinite point ${z\rightarrow\infty}$, and ${F_1=0}$ is
for getting the real solution of $F(z)$. Hence, from
Eqs.~(\ref{eq:ms-orevline-C}) and (\ref{eq:ms-F})
one can obtain that, with ${V\geqslant V_D}$, the coefficient ${b=0}$
in Eq.~(\ref{eq:ms-C_phi}).

For obtaining $F_0$ it is necessary that Eq.~(\ref{eq:ms-C(x,z)}),
after substitution of Eqs.~(\ref{eq:ms-orevline-C}) and (\ref{eq:ms-F}),
must satisfy to Eq.~(\ref{eq:ms-C_phi}) on the perturbed interface
(\ref{eq:ms-perturbation}) in the first order of magnitude. It leads
to the following expression
\begin{equation}
  \label{eq:ms-F0}
  F_0 = b - G_C.
\end{equation}
Hence, substituting Eqs.~(\ref{eq:ms-orevline-C}) and
(\ref{eq:ms-F})--\eqref{eq:ms-F0} into Eq.~(\ref{eq:ms-C(x,z)}),
one gets an expression for the perturbed field of 
solute concentration. This yields
\begin{equation} \label{eq:ms-C-C_0}
  C - C_0 =
  \begin{cases}
    \begin{split}
      \displaystyle &\frac{G_C D (1-V^2/V_D^2)}{V}
      \left[
        1 - \exp \left(
          -\frac{Vz}{D(1-V^2/V_D^2)}
        \right)
      \right] \\  %%%[12pt]
      &+ (b-G_C)\delta (t) \sin(\omega x)
      \exp\left(
        -\frac{\omega_Cz}{(1-V^2/V_D^2)^{1/2}}
      \right),
    \end{split}
    & V<V_D, \\[24pt]
    0, & V\geqslant V_D.
  \end{cases}
\end{equation}
Within the local equilibrium limit $V_D \rightarrow \infty$
(i.e. when the interface velocity is much smaller than the diffusion speed, ${V \ll
  V_D}$), solution (\ref{eq:ms-C-C_0}) transforms into solution obtained
in Ref.~\cite{Trivedi-Kurz} for the case of the local equilibrium solute diffusion
transport. Furthermore, as Eq.~(\ref{eq:ms-C-C_0}) shows, with the velocities equal 
to or greater than the solute diffusion speed the concentration does not depend 
on the interfacial perturbations and is equal to those one at the planar interface. 
It is known from solution given in Ref. \cite{GS} that the solute concentration at 
the unperturbed planar interface is described by 
\begin{equation} \label{ubi109}
  C (z) - C_\infty =
  \begin{cases}
    \begin{split}
      \displaystyle &\frac{1-k}{k}C_\infty \exp \bigg( - \frac {Vz}{D(1-V^2/V_D^2)} \bigg),
    \end{split}
    & V<V_D, \\[24pt]
    0, & V\geqslant V_D, 
  \end{cases}
\end{equation}
where $C_{\infty}$ is the solute concentration in the liquid far from
the interface (i.e., nominal concentration of the alloy). Then, from 
Eqs.~(\ref{eq:ms-C-C_0}) and (\ref{ubi109}) it is clear that 
\begin{equation} \label{ubi110}
  C (x,z) =C_0 \equiv C_\infty \ with \ V\geqslant V_D.  
\end{equation}
Therefore, even in the presence of perturbations, a transition 
to the complete partitionless solidification, $C(x,z)=C_\infty$, proceeds 
in the alloy with the finite velocities $V\geqslant V_D$. 

A solution of Eqs.~(\ref{eq:ms-steady-state-T_L}) and
(\ref{eq:ms-steady-state-T_S}) for the heat transfer with the condition (\ref{eq:ms-T_phi})
is obtained in Ref.~\cite{Trivedi-Kurz} and has the following form
\begin{equation}
  \label{eq:ms-T_L-T_0}
  \begin{split}
    T_L-T_0 & = \frac{G_L a_L}{V}
    \left[
      1 - \exp \left(
        -\frac{Vz}{a_L}
      \right)
    \right] \\
    & + (a-G_L)\delta (t) \sin(\omega x) \exp(-\omega_L z),
  \end{split}
\end{equation}
\begin{equation}
  \label{eq:ms-T_S-T_0}
  \begin{split}
    T_S-T_0 & = \frac{G_S a_S}{V}
    \left[
      1 - \exp \left(
        -\frac{Vz}{a_S}
      \right)
    \right] \\
    & + (a-G_S)\delta (t) \sin(\omega x) \exp(\omega_S z),
  \end{split}
\end{equation}
where $G_L$ and $G_S$~are the gradients of temperature in the liquid and solid
on the unperturbed planar interface, respectively. 
The frequencies $\omega_L$ and $\omega_C$ are described by
\begin{equation}
  \label{eq:ms-omega_L}
  \omega_L = \frac{V}{2a_L}
  + \left[
    \left(
      \frac{V}{2a_L}
    \right)^2
    + \omega^2
  \right]^{1/2},
\end{equation}
\begin{equation}
  \label{eq:ms-omega_S}
  \omega_S = - \frac{V}{2a_S}
  + \left[
    \left(
      \frac{V}{2a_S}
    \right)^2
    + \omega^2
  \right]^{1/2}.
\end{equation}

Using the transport balances (\ref{(2t)}) and (\ref{(2)}), 
the boundary conditions on the perturbed interface $\phi(x,t)$,
Eq.~(\ref{eq:ms-perturbation}), are obtained for 
the steady-state regime of solidification as follows 
\begin{equation}
  - K_L
  \left.
    \frac{\partial T_L}{\partial z}
  \right|_\phi
  + K_S
  \left.
    \frac{\partial T_S}{\partial z}
  \right|_\phi
  = Q \widetilde{V},
  \label{eq:ms-thermal-balance}
\end{equation}
\begin{equation}
  -D (1-\widetilde{V}^2/V_D^2)
  \left.
    \frac{\partial C}{\partial z}
  \right|_\phi
  \left.
    = (1-k)\widetilde{V}C
  \right|_\phi .
  \label{eq:ms-concentration-balance}
\end{equation}
Here $\widetilde{V}$~is the velocity of the perturbed interface. 
Substitution of the expression $\widetilde{V}=V+ (d\delta /dt) \sin (\omega x)$ for the
velocity of the perturbed interface into
Eqs.~(\ref{eq:ms-thermal-balance}) and
(\ref{eq:ms-concentration-balance}) gives the condition of stability
regarding the sign of the function $\delta^{-1} d\delta /dt$. 
The concrete form of the function $\delta^{-1} d\delta /dt$ is obtained as follows.

In the zero order of magnitude by the amplitude $\delta(t)$ of
perturbation, substitution of
Eqs.~(\ref{eq:ms-K})--(\ref{eq:ms-T_phi}) into Eq.~(\ref{eq:ms-Tf})
gives the following relation
\begin{equation}
  \label{eq:ms-T0-C0}
  T_0 = T_m + m C_0,
\end{equation}
for the temperature $T_0$ and solute concentration $C_0$ on the unperturbed
planar interface. Eq.~(\ref{eq:ms-T0-C0}) is also consistent with the liquidus 
line in the kinetic diagram of phase state.
In the first order of magnitude by the amplitude $\delta(t)$, one gets
a relation between coefficients $a$ and $b$ in Eqs.~(\ref{eq:ms-C_phi}) and
(\ref{eq:ms-T_phi}). This yields
\begin{equation}
  \label{eq:ms-a-b}
  a = mb - \Gamma \omega^2.
\end{equation}
Equation (\ref{eq:ms-a-b}) gives a linear relation for the corrections to the temperature
and solute concentration on the perturbed interface and uses Gibbs-Thomson effect for
the curved interface. For the heat balance at the interface, let us now
substitute $\widetilde{V}=V + (d\delta /dt) \sin (\omega x)$ for
the interface velocity into the condition (\ref{eq:ms-thermal-balance}).
Then, in the zero order of magnitude by the amplitude $\delta(t)$,
one gets a relation for the temperature gradients and the velocity
of the unperturbed planar interface. This is
\begin{equation}
  \label{eq:ms-GL-GS-V}
  - K_L G_L + K_S G_S = QV.
\end{equation}
In the first order of magnitude one gets the following expression
\begin{equation}
  \label{eq:ms-GL-GS-V-1}
  Q \frac{1}{\delta} \frac{d\delta}{dt}
  = a(K_L\omega_L+K_S\omega_S)
  + K_L G_L( V/a_L-\omega_L )
  - K_S G_S( V/a_S+\omega_S ),
\end{equation}
which gives the change of the amplitude $\delta(t)$ of perturbation in
time according to the heat balance. Substituting expression
$\widetilde{V}=V+ (d\delta/dt) \sin (\omega x)$ into the mass balance
(\ref{eq:ms-concentration-balance}), one can get in the zero order of
magnitude a relation for the gradient of concentration, interface
velocity and solute concentration in the following form
\begin{equation}
  \label{eq:ms-GC-C0-V}
  -D (1-V^2/V_D^2)G_C = (1-k)VC_0.
\end{equation}
According to Eq.~(\ref{eq:ms-concentration-balance}),
the change of the amplitude $\delta(t)$ of perturbation in time
is defined by the first order of smallness by the amplitude
of perturbation. This yields
\begin{equation}
  \label{eq:ms-GC-C0-V-1}
  \begin{split}
    (C_0 - 2DG_CV/V_D^2)\frac{1}{\delta}\frac{d\delta}{dt}
    & =  b \left[
      D \omega_C (1-V^2/V_D^2)^{1/2}
      - (1-k)V
    \right] \\
    & + G_C \left[
      V - D\omega_C (1-V^2/V_D^2)^{1/2}
    \right].
  \end{split}
\end{equation} 
Note that due to introduction of the finite speed $V_D$ of solute
   diffusion into the model, an additional term $-2DG_CV/V_D^2$ has
   appeared in the left-hand side of Eq.~(\ref{eq:ms-GC-C0-V-1}) in
   comparison to the analysis in \cite{Trivedi-Kurz}.  The prefactor
   $(C_0 - 2DG_CV/V_D^2)$ has a positively defined value: one gets
   $(C_0 - 2DG_CV/V_D^2)>0$ due to negative concentrational gradient
   $G_C \leqslant 0$ with $k<1$ at any velocity $V<V_D$.  Then, to
   obtain limiting cases, one can take the expression for $G_C$ from
   the balance of solute concentration (\ref{eq:ms-GC-C0-V}) in the
   form $G_C = - (1-k) V C_0/(D (1-V^2/V_D^2))$, and one gets the
   relation $(C_0 - 2DG_CV/V_D^2)=C_0(1 +
   2(1-k)(V^2/V_D^2)/(1-V^2/V_D^2))$. From this it follows the {\it
     first positive asymptotic}: $(C_0 - 2DG_CV/V_D^2)=C_0 > 0$ with
   $V_D \rightarrow \infty$. For obtaining the second asymptotic with
   $V \rightarrow V_D$, let us consider the interface velocity in the
   vicinity of the diffusion speed, i.e. the velocity
   $V=V_D-\varepsilon$, $0 \leqslant \varepsilon \ll 1$.  Then, for $V
   \rightarrow V_D$, one can obtain the prefactor in the form
   \begin{displaymath}
     \begin{split}
       & \bigg( C_0 - \frac{2DG_CV}{V_D^2} \bigg)
       \bigg|_{V=V_D-\varepsilon}
       = C_0 \bigg(
       1 + \frac{2(1-k(V))V^2/V_D^2}{1-V^2/V_D^2}
       \bigg) \bigg|_{V=V_D-\varepsilon} \\
       & = C_0 \bigg( 1 + (1-k(V_D))\frac{V_D}{\varepsilon}
       + V_D\frac{dk(V)}{dV}\bigg|_{V=V_D} -2(1-k(V_D)) \bigg).
     \end{split}
   \end{displaymath}
   From this expression follows two consequences. First, in the case
   of partition solidification with $V=V_D$, one gets $k(V_D)\neq 1$.
   In this case, the prefactor tends to infinity with $\varepsilon
   \rightarrow 0$.  Second, with the complete solute trapping,
   $k(V_D)=1$ and $dk(V)/dV \geqslant 0$ one gets for $V \rightarrow
   V_D$ that the prefactor has {\it a positive sign and limited
     magnitude for the second asymptotic}.  This yields
   \begin{displaymath}
     0 < C_0 \bigg( 1 + \frac{2(1-k(V))V^2/V_D^2}{1-V^2/V_D^2} \bigg) \\
     = C_0 \bigg( 1 + V_D\frac{dk(V)}{dV}\bigg|_{V=V_D} \bigg) < \infty.
   \end{displaymath}
   Consequently, $(C_0 - 2DG_CV/V_D^2)$ is positively defined also for
   the local equilibrium case $V_D \rightarrow \infty$ and local
   nonequilibrium case $V \rightarrow V_D$ if the complete solute
   trapping occurs, $k(V_D)=1$, and influences only the speed of
   decreasing/increasing of the amplitude of perturbation, but not a
   selection of the stable mode itself.

From equations (\ref{eq:ms-GL-GS-V-1}) and (\ref{eq:ms-GC-C0-V-1}), 
taking into account Eq.~(\ref{eq:ms-a-b}), the expression for the 
function $\delta^{-1} d\delta / dt$ can be obtained. 
The sign of $\delta^{-1} d\delta / dt$
defines the condition  of decreasing, $\delta^{-1} d\delta / dt<0$,
or increasing, $\delta^{-1} d\delta / dt>0$, of the interfacial perturbation
in time. With $\delta^{-1} d\delta / dt=0$ one has the marginal (neutral)
stability of the interface \cite{Mullins-Sekerka-1964,Trivedi-Kurz}.

%%% MARGINAL STABILITY %%%%%%%%%%%%%%%%%%%%%%%%%%%%%%%%%%%%
\section{Marginal stability} 
\subsection{The criterion of marginal stability}
We now consider marginal stability for the neutral stability
of a small perturbation on the planar interface, 
$\delta^{-1}d\delta / dt=0$.
From equations (\ref{eq:ms-GL-GS-V-1}) and (\ref{eq:ms-GC-C0-V-1}), 
one can obtain expressions for the values of $a$ and $b$. These are
\begin{equation}
  \label{eq:ms-ms-a}
  a = K_L G_L \frac{\omega_L - V/a_L}{K_L\omega_L + K_S\omega_S}
  + K_S G_S \frac{\omega_S - V/a_S}{K_L\omega_L + K_S\omega_S},
\end{equation}
\begin{equation}
  \label{eq:ms-ms-b}
  b = \begin{cases}
    G_C \displaystyle\frac{\omega_C - V/[ D (1-V^2/V_D^2)^{1/2} ]}
    {\omega_C - (1-k)V/[ D (1-V^2/V_D^2)^{1/2} ]},
    & V<V_D, \\ %[24pt]
    0, & V\geqslant V_D.
  \end{cases}
\end{equation}
The system of equations (\ref{eq:ms-a-b}), (\ref{eq:ms-ms-a}), and
(\ref{eq:ms-ms-b}) allows one to obtain a relation for the constant front velocity $V$
and the frequency $\omega$ of a perturbation in the steady-state regime by
excluding of $a$ and $b$. This relation can be considered
as a final form for the condition of marginal stability.

Let's introduce the following functions of stability
\begin{equation}
  \label{eq:ms-ms-xi_L}
  \xi_L = \frac{\omega_L - V/a_L}{K_L\omega_L + K_S\omega_S},
\end{equation}
\begin{equation}
  \label{eq:ms-ms-xi_S}
  \xi_S = \frac{\omega_S + V/a_S}{K_L\omega_L + K_S\omega_S},
\end{equation}
\begin{equation}
  \label{eq:ms-ms-xi_C}
  \xi_C = \begin{cases}
    \displaystyle
    \frac{\omega_C - V/[ D (1-V^2/V_D^2)^{1/2} ]}
    {\omega_C - (1-k)V/[ D (1-V^2/V_D^2)^{1/2} ]},
    & V<V_D, \\ %[24pt]
    0, & V\geqslant V_D.
  \end{cases}
\end{equation}
The functions $\xi_L$ and $\xi_S$ in Eqs.~(\ref{eq:ms-ms-xi_L})
and (\ref{eq:ms-ms-xi_S}) coincide with those derived 
by Trivedi and Kurz \cite{Trivedi-Kurz}. 
However, as Eq.~(\ref{eq:ms-ms-xi_C})
shows, the function $\xi_C$ of concentrational
stability differs from the corresponding function derived previously in
Ref.~\cite{Trivedi-Kurz}. As it follows from Eq.~(\ref{eq:ms-ms-xi_C}),
within the local equilibrium limit, $V_D \rightarrow \infty$, one gets
the special case $\xi_C=(\omega_C-V/D)/(\omega_C-(1-k)V/D)$ which
coincides with the result given in Ref.~\cite{Trivedi-Kurz}.
When ${V \sim V_D}$, the function $\xi_C$ given by Eq.~(\ref{eq:ms-ms-xi_C})
takes corrections for the relation of the interface velocity, $V$, and diffusion
speed, $V_D$. With ${V\geqslant V_D}$, the exact equality ${\xi_C=0}$
takes place. This equality is the consequence of solution of the problem of
local nonequilibrium solute diffusion which takes into account
the finite speed $V_D$ in the bulk liquid. Thus, after substituting 
Eqs.~(\ref{eq:ms-ms-a})
and (\ref{eq:ms-ms-b}) into Eq.~(\ref{eq:ms-a-b}) and taking into
account Eqs.~(\ref{eq:ms-ms-xi_L})--(\ref{eq:ms-ms-xi_C}), one can obtain
the criterion of marginal stability. This yields 
\begin{equation}
  \label{eq:ms-marginal-stability-criterion}
  \begin{cases}
    \Gamma \omega^2 + K_L G_L \xi_L + K_S G_S \xi_S - m G_C \xi_C =0,
    & V<V_D, \\ %[12pt]
    \Gamma \omega^2 + K_L G_L \xi_L + K_S G_S \xi_S =0,
    & V\geqslant V_D.
  \end{cases}
\end{equation}
In the local equilibrium limit, $V_D \rightarrow \infty$,
criterion (\ref{eq:ms-marginal-stability-criterion}) transfers into the criterion
of marginal stability obtained in Ref.~\cite{Trivedi-Kurz} on the basis of a 
local equilibrium approach to solute diffusion transport.
The introduction of the finite diffusion speed, $V_D$, into the model leads to
the qualitatively new result, which is related to the transition to completely
partitionless solidification.
As Eq.~(\ref{eq:ms-marginal-stability-criterion}) shows, with the finite
interface velocity ${V\geqslant V_D}$, the solute diffusion ahead of the rapid 
interface is absent [see solution (\ref{eq:ms-C-C_0})], 
and the morphological stability is defined by the relation of the stabilizing
force ${\Gamma\omega^2}$, due to surface energy, and the
contribution $K_L G_L \xi_L + K_S G_S \xi_S$ of 
temperature gradients, $G_L$ and $G_S$.

Using criterion (\ref{eq:ms-marginal-stability-criterion}), one can 
analyze qualitatively two different situations for solidification
when (i) the latent heat is removed from the interface inside the
undercooled liquid phase, and (ii) the latent heat is removed from
the interface through the solid crystal phase.  In case (i), one gets
${K_L G_L \xi_L} + {K_S G_S \xi_S} < 0$, and the temperature gradient
is destabilizing the interface.  Therefore, if the absolute morphological 
stability is not reached by the steady balance 
${\Gamma \omega^2}$ = ${K_L G_L \xi_L} + {K_S G_S \xi_S}$, 
the interface is unstable, and the resulting interface may
exhibit a cellular-dendritic pattern.  In case (ii), with ${K_L G_L
  \xi_L} + {K_S G_S \xi_S} > 0$, the total heat flux is directed from
the front to the solid phase and the temperature gradient, in
addition to the surface energy, stabilizes the form of the solid-liquid
interface. 
In this case, with ${V<V_D}$, the morphological stability is related to 
a destabilizing  action of the force ${m G_C   \xi_C}$, 
directly connected with the concentrational gradient $G_C$, and the 
stabilizing force 
${\Gamma \omega^2} + {K_L G_L \xi_L} + {K_S G_S  \xi_S}$, 
due to the sum of the total positive temperature gradient and the 
surface energy. With the interface velocity ${V\geqslant  V_D}$, 
solidification leads to the chemically partitionless pattern. 
A destabilizing action on the front is absent and the interface 
itself remains linearly stable against any small interfacial perturbation.

\subsection{Characteristic size for crystal microstructure}
According to the marginal stability hypothesis suggested in Ref.~\cite{LMK} 
and developed
in Refs.~\cite{Kurz-Fisher,Trivedi-Kurz-2}, a characteristic size $R$
selected by crystal microstructure  in solidification (e.g., the dendrite 
tip radius) is related to the critical wavelength, $\lambda$, of interface
perturbation as
\begin{equation}
  \label{eq:ms-marginal-stability-hypothesis}
  R \equiv \lambda = \frac{\omega}{2\pi}.
\end{equation}
Assuming equality for thermophysical parameters of the liquid and solid,
one can obtain characteristic size, $R$,
from Eqs.~(\ref{eq:ms-marginal-stability-criterion}) and
\eqref{eq:ms-marginal-stability-hypothesis}. This yields \\
{\it - with $V<V_D$}
\begin{equation}
  \label{eq:ms-R-V<VD}
  R =
  \left(
    \frac{\Gamma/\sigma}{
      m G_C \bar\xi_C
      - \frac{1}{2} (G_L\bar\xi_L + G_S\bar\xi_S)
      }
  \right)^{1/2},
\end{equation}
{\it - with $V\geqslant V_D$}
\begin{equation}
  \label{eq:ms-R-V>VD}
  R =
  \left(
    \frac{\Gamma/\sigma}{- \frac{1}{2} (G_L\bar\xi_L + G_S\bar\xi_S)}
  \right)^{1/2}.
\end{equation}
In Eqs.~\eqref{eq:ms-R-V<VD} and  \eqref{eq:ms-R-V>VD} the following
designations are accepted
\begin{equation}
  \label{eq:ms-sigma}
  \sigma = \frac{1}{4\pi^2}, 
\end{equation}
\begin{equation}
  \label{eq:ms-bar-xi_L}
  \bar\xi_L = 1 - \frac{1}{
    \left( 1 + \displaystyle \frac{1}{\sigma P_T^2} \right)^{1/2}
},
\end{equation}
\begin{equation}
  \label{eq:ms-bar-xi_S}
  \bar\xi_S = 1 + \frac{1}{
    \left( 1 + \displaystyle \frac{1}{\sigma P_T^2}\right)^{1/2}
    },
\end{equation}
\begin{equation}
  \label{eq:ms-bar-xi_C}
  \bar\xi_C = 1 + \frac{2k}{
    1 - 2k
    - \left( 1 + \displaystyle \frac{1-V^2/V_D^2}{\sigma P_C^2}\right)^{1/2}
    },
\end{equation}
with $P_T=VR/2a$ and $P_C=VR/2D$ as the thermal and
solutal Peclet numbers, respectively.
As we noted above, Eq.~\eqref{eq:ms-R-V>VD} is true only for solidification in an 
undercooled liquid, i.e. when the temperature gradient is negative. 
For the case of solidification in the positive temperature gradient 
the absolute morphological stability takes place at the interface velocity, $V$, 
smaller than the diffusion speed, $V_D$, in the liquid.

%%% ABSOLUTE STABILITY %%%%%%%%%%%%%%%%%%%%%%%%%%%%%%%%
\section{Absolute stability}
\subsection{Nonisothermal solidification}
Let's consider equation (\ref{eq:ms-marginal-stability-criterion}) in the limit
of large perturbations wavelengths ${\lambda\gg 1}$, which is true for 
${\omega\ll 1}$. With this condition, from Eqs.~(\ref{eq:ms-ms-xi_L}) and
(\ref{eq:ms-ms-xi_C}) one can yield expansions for the functions $\xi_L$ and
$\xi_C$ in the following form
\begin{equation}
  \label{eq:ms-xi_L-xi_C-omega}
  \xi_L = \frac{a_L^2 \omega^2}{K_L V^2},
  \qquad
 \xi_C = \begin{cases}	
    \displaystyle
    \frac{\omega^2 D^2 (1-V^2/V_D^2)}{k V^2},
    & V<V_D, \\ %[24pt]
    0, & V\geqslant V_D.
\end{cases}
\end{equation}
Substituting these expressions into the criterion of marginal stability
(\ref{eq:ms-marginal-stability-criterion}) at $G_S=0$, we get the criterion
of absolute morphological stability for the planar front. This criterion 
can be written in the form of the following nonlinear equation for the 
velocity, $V_A$, of absolute stability: 
\begin{equation}
  \label{eq:ms-VA(GL)}
  V_A = V^T_A(V) + V^C_A(V), 
\end{equation}
where
\begin{equation}
  \label{eq:ms-VA_T}
V^T_A(V) = 
 - \frac{a_L}{\Gamma} \left(\frac{a_L G_L}{V}\right) 
\end{equation} 
is the velocity of absolute thermal stability, and 
\begin{equation}
  \label{eq:ms-VA_C}
V^C_A(V) = 
    \frac{D}{\Gamma k}\left(\frac{D(1-V^2/V_D^2)mG_C}{V}\right)<V_D 
\end{equation} 
is the velocity of absolute chemical stability. 

The velocity, $V^T_A$, of absolute thermal stability, 
Eq.~\eqref{eq:ms-VA_T}, shows the relationship between the contribution of the 
temperature gradient, $G_T$, and capillary parameter, $\Gamma$. 
With the negative temperature gradient, $G_L<0$, the range of morphological  
stability shrinks. Conversely, the positive temperature gradient, $G_L>0$, extends the 
range of the velocities at which the planar front is linearly stable. 
The velocity, $V^C_A$, of absolute chemical stability, Eq.~\eqref{eq:ms-VA_C}, 
defines the contribution of the concentrational 
gradient, $G_C$, and capillary parameter $\Gamma$. 
This velocity is always less than the diffusion 
speed, $V^C_A<V_D$, because the velocity $V^C_A$ is defined by the steady 
balance between surface tension, given by the capillary parameter $\Gamma$, 
and the gradient $G_C$ of solute concentration, existing 
up to the completion of solute diffusion. 
Consequently, Eqs.~\eqref{eq:ms-VA(GL)}-\eqref{eq:ms-VA_C} 
exhibit a competition of destabilizing and stabilizing forces. With
the velocity, $V<V_A$, the planar interface is perturbed with a 
possible originating of the 
cellular-dendritic patterns. As the solidification velocity increases, 
$V>V_A$, the planar interface becomes morphologically stable against 
any small perturbation of its form. 

To clarify contributions from both thermal  
and solute diffusion on the absolute stability of the interface,
we define the gradients 
in Eqs.~(\ref{eq:ms-VA_T})-(\ref{eq:ms-VA_C}) 
in explicit form. For the thermal and concentrational gradients 
at the unperturbed planar interface we use a solution of the 
local-nonequilibrium problem \cite{GS}. From the solution, 
one gets 
\begin{equation}
  \label{eq:ms-VA-GC}
G_L = -\frac{T_QV}{a_L}, \qquad
  G_C =
  \begin{cases}
    - \displaystyle \frac{(1-k)VC_{\infty}}{kD(1-V^2/V_D^2)},
    & V<V_D, \\ %[24pt]
    0, & V \geqslant V_D,
  \end{cases}
\end{equation}
where $T_Q$ is a unit of undercooling equal to $Q/\chi_L$, and 
$C_{\infty}$ is the solute concentration in the liquid far from
the interface. From the second expression 
in Eq.~(\ref{eq:ms-VA-GC}), it follows that when the solute diffusion 
is absent ahead of the interface with $V \geq V_D$, the gradient of 
the solute concentration is zero exactly. 
Substitution of Eq.~(\ref{eq:ms-VA-GC}) into 
Eqs.~(\ref{eq:ms-VA(GL)})-(\ref{eq:ms-VA_C}) 
gives the expression for the absolute stability of the interface. This yields 
\begin{equation}
  \label{VATC}
  V_A =\begin{cases}
    \displaystyle V^T_A + V^C_A,
    & V<V_D, \\ \\
    \displaystyle V^T_A, & V \geqslant V_D,
  \end{cases}
=
  \begin{cases}
    \displaystyle \frac{a_L}{\Gamma} \Delta T_T + \frac{D}{\Gamma k} \Delta T_C,
    & V<V_D, \\ \\
    \displaystyle \frac{a_L}{\Gamma} \Delta T_T, & V \geqslant V_D,
  \end{cases}
\end{equation}
where $\Delta T_T=T_Q$ is the thermal undercooling, 
which is necessary for solidification with the planar 
interface on the thermal scale, and $\Delta T_C=(k-1)mC_\infty /k$ 
is the constitutional undercooling, which is necessary for 
solidification with the planar interface on the scale of solute diffusion. 
Additionally, $\Delta T_C$ is the nonequilibrium temperature 
interval of solidification between liquidus and solidus lines in 
the kinetic diagram of phase state. 

The criterion $V^T_A=a_L\Delta T_T/\Gamma$ in Eq.~(\ref{VATC}) 
is the same as that which has been obtained by Trivedi and Kurz \cite{Trivedi-Kurz} 
using the advanced model for large growth velocities. 
The criterion $V^C_A=D\Delta T_C/\Gamma=D(k-1)mC_\infty /(\Gamma k^2)$ 
in Eq.~(\ref{VATC}) is similar to that which has been obtained 
by Mullins and Sekerka \cite{Mullins-Sekerka-1964} 
for the case of small growth velocities, and re-derived by Trivedi and Kurz 
\cite{Trivedi-Kurz} for the case of rapid solidification. 
In addition to this treatment, by 
introducing the finite speed, $V_D$, into the model 
we reach a qualitative new result. At the finite velocity, $V \geq V_D$, 
due to the absence of the solute diffusion 
[$G_C=0$, Eq.~(\ref{eq:ms-VA-GC})], the interval between nonequilibrium 
liquidus and solidus lines is equal to zero, $\Delta T_C=0$. 
These lines converge in the kinetic phase diagram with $V \geq V_D$ \cite{GS}, 
and the absolute stability of the planar interface 
is defined only by the undercooling $\Delta T_T$, and relation 
between the thermal diffusivity, $a_L$, and capillary parameter, $\Gamma$, 
Eq.~(\ref{VATC}).  

\subsection{Isothermal solidification}
In the analysis of the criterion of marginal stability
(\ref{eq:ms-marginal-stability-criterion}), a special interest is given to the case
in which the form of the interface is defined by the competition between
stabilizing force, ${\Gamma \omega^2}$, due to surface energy, 
and destabilizing force, ${m G_C \xi_C}$, due to gradient, $G_C$, of
solute concentration. Assuming the zero
temperature gradient, $G_L=0$, in Eq.~(\ref{eq:ms-VA(GL)}), 
one can obtain an explicit expression for the 
condition of absolute chemical stability of the interface, $V_A=V_A^C$.
Using the expression for $G_C$ from (\ref{eq:ms-VA-GC}), one gets
\begin{equation}
  \label{eq:ms-VA}
  V_A = \frac{mD(k-1)C_{\infty}}{\Gamma k^2} < V_D.
\end{equation}
The form of this expression coincides with the expression 
given for the case of local equilibrium solute diffusion 
transport at $V_D \rightarrow \infty$ and 
${V \ll V_D}$ \cite{Trivedi-Kurz}. However, a final form 
of the function $V_A(C_{\infty})$ is defined by the functions 
of solute partitioning, $k(V)$, and the slope, $m(V)$, of 
liquidus line in the kinetic phase diagram. The behavior 
of theses functions is rather different for the cases of local equilibrium 
and local nonequilibrium solute diffusion \cite{GS,G-00}.

\section{Discussion} 
In the first part of the discussion, we synthesize 
our system of equations to give the self-consistent 
model, which is adopting  the deviation from 
local equilibrium in the solute diffusion field for all functions. 
We discuss the solute partitioning functions 
and the expression for the slope of kinetic 
liquidus which take into account the deviation from 
local equilibrium, both at the interface and in the bulk 
liquid around the interface. Then, in the second part of the 
discussion, we present a quantitative evaluation of the 
discrepancies between the present model and the 
model in which the local nonequilibrium is taken only at 
the interface. These are compared with 
experimental data on the absolute stability 
of the planar interface. 

\subsection{Solute partitioning and kinetic liquidus}
The boundary condition for solute diffusive transport can be
given on the basis of the continuous growth model (CGM) \cite{Az-Kap'88}. The
CGM gives the solute partitioning function at the solid-liquid interface, which, in
the dilute solution approximation, is described by \cite{Az-Kap'88,az03}
\begin{equation}
k(V)= \frac{k_e+V/V_{DI}}{1+V/V_{DI}},
\label{28}
\end{equation}
where $V_{DI}$ is the speed of diffusion at the interface, 
and $k_e$ is the value of the equilibrium partition coefficient with 
$V \rightarrow 0$. 
One of the deficiencies of the function (\ref{28}) is the difficulty
to describe of the complete solute trapping
at the finite interface velocity,
i.e. it predicts $k\rightarrow 1$ only with $V\rightarrow \infty$.
However, as it has been shown in numerous experiments (see,
e.g., Refs.~\cite{mir,eck001}), a transition to partitionless solidification
occurs at a finite solidification velocity. Furthermore, the molecular
dynamic simulation has shown \cite{cook} that the transition
to the complete solute trapping is observed at finite
crystal growth velocity. Therefore, in addition to Eq.~(\ref{28}),
a generalized function for solute partitioning, in the case of local
nonequilibrium solute diffusion within the approximation of a dilute
alloy, has been introduced \cite{S1}. This yields
\begin{equation}
k(V) = \begin{cases}	
    \displaystyle
    \frac{k_e(1-V^2/V_D^2)+V/V_{DI}}{1-V^2/V_D^2+V/V_{DI}},
    & V<V_D, \\ %[24pt]
    1, & V\geqslant V_D, 
  \label{25}
\end{cases}
\end{equation}
where $V_{DI}$ is the interfacial diffusion speed with
$V_{DI}\leq V_D$ \cite{GD1,S1}. 
In the local equilibrium limit, i.e. when the bulk
diffusive speed is infinite, $V_D \rightarrow \infty$,
expression (\ref{25}) reduces to the function $k(V)$, that takes
into account the deviation from local equilibrium at the interface
only, Eq.~(\ref{28}). In addition to the previous model 
\cite{Az-Kap'88,az03}, the function $k(V)$ described by Eq. (\ref{25})
includes the deviation from local equilibrium not only
at the interface (introducing interfacial diffusion speed $V_{DI}$), but also
in the bulk liquid (introducing diffusive speed $V_D$ in bulk). 
As Eq. (\ref{25}) shows, the complete solute trapping,  $k(V)=1$,
proceeds at $V=V_D$.

A thermodynamic approach applied to the solidification of a binary system
\cite{Bak-Cah71} provided two models for the
solute trapping with and without solute drag
\cite{Bo-Co86,azboet}. These
models give a shift from local equilibrium at the interface which can
be expressed in unified form for the slope, $m(V)$, of
kinetic liquidus by the following equation
\begin{equation}
m(V) = \frac {m_e}{1 - k_e}
\bigg\{1 - k + \left [k + (1-k)\delta_0\right]ln \left (\frac {k}{k_e} \right)\bigg\}.
\label{26}
\end{equation}
Here $\delta_0=0$ is for the model of solute trapping without solute drag 
and 
$\delta_0=1$ is for the model of solute trapping with solute drag.
Introducing Eq.~(\ref{28}) into Eq.~(\ref{26}), 
one obtains the constant liquidus slope, $m$, 
(independent of $V$) only with the infinite interface velocity,
$V \rightarrow \infty$.

Using the results of the local nonequilibrium thermodynamic analysis
\cite{G-00}, one arrives to the slope of the liquidus line in the following form
\begin{eqnarray}
m(V) = \begin{cases}	
    \displaystyle
   \frac {m_e}{1 - k_e}
		\bigg\{1 - k + ln \left (\frac {k}{k_e} \right) + (1 - k)^2 \frac {V}{V_D} \bigg\},
    & V<V_D, \\ \\ 
\displaystyle   
 \frac {m_e lnk_e}{k_e- 1}, & V\geqslant V_D. 
  \label{27}
\end{cases}
\end{eqnarray}
With $V<V_D$, the function $m(V)$ includes the function 
described by Eq.~(\ref{26}) for the solute trapping
with solute drag ($\delta_0=1$) and the additional term $(1 - k)^2 V/V_D$. 
This term arises from the analysis of the Gibbs free energy, taking into account
local nonequilibrium solute diffusion around the interface.
It is necessary to note that the function $m(V)$ described
by Eq.~(\ref{27}) plays a 
crucial role for self-consistency of the theory of local nonequilibrium  solidification.
This form of the function has been used 
in a self-consistent model for rapid dendritic growth and gave quantitative
agreement with experimental data on kinetics of alloy solidification
\cite{GD1,GD-99}. In particular, the self-consistent dendritic growth
model, including Eq.~(\ref{27}), predicts the breakpoint at $V=V_D$ with 
good agreement of data on a number of investigated alloys. Furthermore, Eq.~(\ref{27})
gives us the ability to describe a transition from the growth kinetics, with solute drag effect 
at small and moderate solidification velocities (arising with the developed solute
profile ahead of the interface), to the growth kinetics without solute drag at high
solidification velocities (with the degeneration of the solute profile 
ahead the interface) \cite{G-00,G-MSE}. Thus, Eqs.~(\ref{25}) and (\ref{27}) 
close the system of equations (\ref{eq:ms-steady-state-C})-(\ref{eq:ms-K}) 
for the self-consistent analysis of morphological stability. 

\subsection{Comparison with experimental data}
To discuss the results obtained for the interfacial stability,
we now compare the model predictions for the absolute stability condition 
(\ref{eq:ms-VA}) in both 
cases of solidification, namely, with local equilibrium solute diffusion and with
local nonequilibrium solute diffusion transports. Substituting
functions (\ref{28})-(\ref{27}) into Eq.~(\ref{eq:ms-VA}), we analyze the absolute
stability of the planar interface for different velocities. 
We stress two important points regarding the choice of the expression for 
the slope of kinetic liquidus 
given by Eq.~(\ref{26}). {\it First}, the result on rapid dendritic growth 
\cite{GD1,GD-99} gives evidence to the confluence of all model predictions at small 
undercoolings and low growth velocities. Disagreement of the kinetic curves 
begins from the undercooling approximately corresponding to 
the undercooling for the absolute chemical stability at moderate growth velocities. 
Therefore, our present discussion for interfacial stability is limited by 
the moderate interface velocities, when the model's predictions 
(with or without taking local nonequilibrium in bulk liquid) begin to disagree. 
{\it Second}, in this region of velocities, the local nonequilibrium approach 
to rapid solidification gives a similar result with the model, which takes into 
account the deviation from local equilibrium only at the interface, with 
the solute-drag effect (see the analysis presented in Refs.~\cite{G-00,G-MSE}). 
This fact is due to the existence 
of the developed solute profile ahead of the interface at small and moderate 
velocities, when the solute drag may appear at the interface. Consequently, 
in order to evaluate the disagreement between the model and experimental data, 
we choose the expression for the slope of the kinetic liquidus, which adopts the solute-drag effect, 
i.e. it is chosen in the following calculations where $\delta_0=1$ for Eq.~(\ref{26}). 

Using parameters of an Al-Fe alloy from Table \ref{tab:1}, one can
calculate the curve for critical concentration, $C_{\infty}(V)$, which
gives a threshold for interface instability. As it can be seen from
Fig.~\ref{fig:ms-VA-C0}, two regions of the interfacial existence
may occur: the planar interface is absolutely stable below the curves
and the interface breaks down in the regions above the curves given by
the functions $C_{\infty}(V)$.  In comparison with the model with the local
equilibrium diffusion and deviation from local equilibrium at the
interface only [Eqs.~(\ref{eq:ms-VA}, (\ref{28}), and (\ref{26})],
the present model for interface stability with the deviation 
from local equilibrium, both at the interface and in the bulk liquid 
[Eqs.~(\ref{eq:ms-VA}), (\ref{25}, and (\ref{27})], 
defines a curve $C_\infty(V)$ which is limited 
by the diffusion speed $V_D$ for morphological stability 
of the interface. This limit exists due to a steady balance 
between the stabilizing capillary force and the destabilizing 
force defined by the concentrational gradient, which 
still acts on the interface up to the finishing of diffusion, 
i.e. until the point $V=V_D$. 

For a quantitative comparison of the model predictions
we have chosen experimental results on interface
stability during rapid solidification of a Si--Sn alloy as
presented by Hoglund and Aziz in Ref.~\cite{H-A}. 
These authors measured a critical concentration of Sn for
interface breakdown in a steady-state 
solidification after pulsed laser melting.
Using the parameters of the Si--Sn alloy from Table \ref{tab:1},  
the model predictions for the function $C_{\infty}(V)$ 
are compared quantitatively with experimental results 
from Ref.~\cite{H-A}. 
As is shown in Fig.~\ref{fig:si_sn_va}, 
the present model for interface stability 
with the local nonequilibrium diffusion 
[Eqs.~(\ref{eq:ms-VA}), (\ref{25}], and (\ref{27})] 
gives a satisfactory comparison with the experiment. 
At the concentration Sn = 0.02 atomic fraction 
(see the extreme right experimental point in Fig.~\ref{fig:si_sn_va}), 
the discrepancy between the model with local equilibrium solute 
diffusion (curve 1 in Fig.~\ref{fig:si_sn_va}) and experiment gives 
the value of 38.90 $\%$ (see Table \ref{tab:2}). 
At the same alloy's concentration, the present model 
(curve 2 in Fig.~\ref{fig:si_sn_va}) gives the discrepancy 
with experiment of 16.93 $\%$ (see Table \ref{tab:2}). 
Consequently, even better comparison with the available 
experimental data can be obtained with using the present 
model of local nonequilibrium solidification. 

%%% CONCLUSIONS %%%%%%%%%%%%%%%%%%%%%%%%%%%%%%%%%%%%%%%%%%%%%%%%%%%%%%%
\section{Conclusions}
Morphological stability of the planar interface  
in rapid solidification of nonisothermal 
binary system has been considered. 
We have taken into account the fact that the high rate of solidification 
process leads to the absence of a local thermodynamic equilibrium 
in the solute diffusion field and at the solid--liquid interface. 
The presently developed model is self-consistent: the main governing 
equations, Sec.~2, and the interface conditions for solute trapping 
and kinetic liquidus, Sec.~6, are consistent with the formalism of extended 
thermodynamic approach to rapid solidification \cite{G-00}. 
Using the model of local nonequilibrium rapid solidification, 
our analysis of morphological stability extends the previous 
analysis of Trivedi and Kurz \cite{Trivedi-Kurz}, which has been 
performed to advance the treatment of Mullins and Sekerka 
\cite{Mullins-Sekerka-1964} to the case of rapid solidification. 
The main outcomes of this analysis are summarized as follows. 

(i) For the velocities equal to or greater than the diffusion speed, 
$V \geq V_D$, from solutions (\ref{eq:ms-C-C_0})-(\ref{ubi110}) 
it follows that the field of concentration does not depend on a form of 
the interfacial perturbation and it is equal to the initial (nominal) concentration, 
$C(x,z)=C_{\infty}$. This result is in agreement with the previous results 
for planar and parabolic interfaces \cite{GS,GD-pla-parab}. 
Solutions (\ref{eq:ms-C-C_0})-(\ref{ubi110}) has a clear physical meaning: 
a source of concentrational disturbances, i.e. the perturbed interface, 
cannot disturb a binary liquid ahead of itself if the interface velocity is equal to 
or greater than the maximum speed of these disturbances.  

(ii) The obtained criterion of the marginal stability, 
Eq.~(\ref{eq:ms-marginal-stability-criterion}), 
defines a wavelength of perturbation for the neutral stability. 
For $V < V_D$, the neutral stability is defined by a balance of 
the stabilizing force, due to surface energy, destabilizing force, 
due to concentrational gradient $G_C$, and the
contribution of temperature gradients $G_L$ and $G_S$. 
Qualitatively, a new result can be obtained from the criterion for the 
front velocity of $V \geq V_D$, i.e. for an absence of the solute diffusion 
ahead of the interface. As it follows from the second equation in 
Eq.~(\ref{eq:ms-marginal-stability-criterion}), the morphological stability 
of the interface is defined only by the relation between the 
thermodynamic stabilizing force, due to the surface tension, 
and the driving force of the morphological instability, 
due to the negative thermal gradient in the undercooled liquid.  
In the case of directional solidification with the positive thermal gradient, 
the destabilizing action on the interface is absent at $V \geq V_D$, 
and the interface remains linearly stable against small perturbations 
of its form. 

(iii) Absolute stability of the planar interface is considered 
as a steady balance between destabilizing force (due to the 
concentrational gradient), the thermal contribution 
(due to the thermal gradient), and the stabilizing force 
(due to surface tension). The velocity, $V_A$, of the absolute 
interface stability is obtained as a sum of the velocity, 
$V^T_A$, for thermal stability and velocity, $V^C_A$, 
for chemical stability defined by Eq.~(\ref{VATC}). 
$V^C_A$ is the same as what was  
obtained by Mullins and Sekerka \cite{Mullins-Sekerka-1964} 
for the case of small growth velocities. It was re-derived by 
Trivedi and Kurz \cite{Trivedi-Kurz} for the case of rapid solidification. 
Introduction of the finite speed, $V_D$, into the model gives 
the qualitatively new result: with the absence of the solute 
diffusion at $V \geq V_D$, the absolute stability of a planar 
interface is defined only by the thermal undercooling 
and relation between the thermal diffusivity, $a_L$, and 
capillary parameter, $\Gamma$, Eq.~(\ref{VATC}). 
For an isothermal solidification, the present analysis 
shows the limiting boundary equals to the diffusion 
speed, $V_D$, for the region of morphological instability 
[see Eq.~(\ref{eq:ms-VA}) and Fig.~\ref{fig:ms-VA-C0}].

(iv) The predictions of the present model 
for the critical concentration above which a planar
interface becomes unstable 
[see Eqs.~(\ref{eq:ms-VA}), (\ref{25}), and (\ref{27})]
are compared with the previous model, which adopts
the deviation from local equilibrium at the interface only
[see Eqs.~(\ref{eq:ms-VA}), (\ref{28}), and (\ref{26})], and  
with the experimental data obtained for solidification of the Si--Sn alloy \cite{H-A}. 
As it is shown in Fig.~ \ref{fig:si_sn_va}, the present model 
is able to describe experimental data satisfactorily in 
a whole region of the interface velocities investigated. 
From numerical evaluation of the theoretical predictions 
summarized in Table \ref{tab:2}, it follows that better 
agreement with experiment is obtained with using 
the present model of local nonequilibrium solidification. 

%%% ACKNOWLEDGMENTS %%%%%%%%%%%%%%%%%%%%%%%%%%%%%%%%%%%%%%
\begin{ack}
This work was performed with support from the DFG Schwerpunktprogramm 1120
under Research Projects He 1601/13 and Ne 822/2-1.
\end{ack}

%%% BIBLIOGRAPHY %%%%%%%%%%%%%%%%%%%%%%%%%%%%%%%%%%%%%%%%%%
\newpage
%References

%%%%%%%   T A B L E   1   %%%%%%%%%%%%%
\def\thetable{\arabic{table}}
\def\thefigure{\arabic{figure}}
%%%%%%%
\newpage
%\begin{small}
\begin{table}[h]
\caption{Physical parameters used in calculations 
of the limit of absolute stability of the planar interface 
in solidifying binary alloys}
\label{tab:1}
\begin{tabular}{llllll}
\hline\hline

Parameter &  &  & Al--Fe & Si--Sn  \\ \hline
Diffusion coefficient & $D_L$ & m$^{2}$/s & $1.7\times10^{-9}~(*)$ & $2.5\times 10^{-8}~(**)$  \\
Partition coefficient & $k_e$ & --- & $0.03~(*)$ & $0.016~(**)$  \\
Liquidus slope & $m_e$ & K/at.\% & $-7.3~(*)$ & $-4.6~(**)$  \\
Gibbs-Thomson coefficient & $\Gamma$ &K $\cdot$ m& $1\times10^{-7}~(*)$
& $1.3\times10^{-7}~(**)$  \\
Interface diffusion speed & $V_{DI}$ & m/s & $7$ & $17~(**)$  \\
Diffusion speed in bulk liquid & $V_D$ & m/s & $10$ & $17.5 $  \\
\hline
(*)~--~Data taken from Ref.~\cite{gremaud}. \\
(**)~--~Data taken from Ref.~\cite{H-A}. \\\\\\\\\\\
\end{tabular}
\end{table} 
%\end{small}

%%%%%%%   T A B L E   2   %%%%%%%%%%%%%
\def\thetable{\arabic{table}}
\def\thefigure{\arabic{figure}}
%%%%%%%
%\newpage
%\begin{small}
\begin{table}[h]
\caption{Discrepancy between theoretical predictions and experiment 
for the absolute chemical stability of the planar interface in the Si--0.02(at.fraction)Sn alloy}
\label{tab:2}
\begin{tabular}{lllll}
\hline\hline
Absolute chemical stability	     &  Velocity & ~~~Definition         & ~~~~~Value  \\
		                             &  ~~(m/s)     & of discrepancy &  of discrepancy (\%) \\ \hline
Local equilibrium solute diffusion,          & & & \\
Eqs.~(\ref{eq:ms-VA}), (\ref{28}), and (\ref{26}) & $V_A^{(1)}=15.5$ 
&  $\displaystyle \frac{V_A^{(1)}-V_A^{(exp)}}{V_A^{(1)}}100 \%$ & ~~~~~~~~~$38.90$ \\\\  \hline 
Local nonequilibrium solute diffusion,          & & & \\
Eqs.~(\ref{eq:ms-VA}), (\ref{25}), and (\ref{27})& $V_A^{(2)}=11.4$ 
&  $\displaystyle \frac{V_A^{(2)}-V_A^{(exp)}}{V_A^{(2)}}100 \%$ & ~~~~~~~~~$16.93$ \\\\  \hline
Experiment, Ref.~\cite{H-A} & $V_A^{(exp)}=9.47$ & ~~~~~~~~--- & ~~~~~~~~~~~---  \\
\hline\hline
\end{tabular}
\end{table}
%\end{small}

\newpage
Figure captions to the article 
``Linear morphological stability analysis 
of the solid-liquid interface
in rapid solidification of a binary system''. \\

\begin{figure}[h]
   \caption{Morphological diagram for solidification of 
     binary systems which is illustrating the microstructural
     transitions ``planar front'' $\rightarrow$ ``cellular structure''
     $\rightarrow$ ``dendrites'' $\rightarrow$ ``cellular
     structure''~$\rightarrow$ ``planar front'', with the increasing of
     the solidification velocity, $V$.  Here $V_C$ is the velocity
     given by the criterion of constitutional undercooling, and $V_A$
     is the velocity for absolute morphological stability of the
     interface. }
\label{fig:morph_spectr-1}
\end{figure}

\begin{figure}[h]
   \caption{Optical micrograph of longitudinal 
through-thickness section of a melt-spun ribbon 
of Ni--18 at.\% B \cite{gal-lad}. 
The crystal microstructure exhibits a transition from 
planar interface with solute segregation-free 
to cellular-dendritic patterns. The transition proceeds due to 
decreasing of the interface velocity from $V>V_A$ 
up to $V<V_A$. Wheel surface at bottom of micrograph.  }
\label{fig:spinning}
\end{figure}

\begin{figure}[h]
  \caption{Velocity $V_A$ of absolute chemical stability versus
    solute concentration $C_{\infty}$ for Al--Fe alloy. 
    Dashed curve corresponds to solution of
    Eqs.~(\ref{eq:ms-VA}), (\ref{28}), and (\ref{26}) with solute-drag effect.  
    Solid curve corresponds to solution of Eqs.~(\ref{eq:ms-VA}), (\ref{25}), and
    (\ref{27}). Dashed-dotted line, $V=V_D$, represents the limiting velocity 
    for the absolute interface stability.  }
\label{fig:ms-VA-C0}
\end{figure}

\begin{figure}[h]
  \caption{Critical concentration, $C_{\infty}$, above which planar
    interface is unstable. Experimental points correspond to 
    solidification of the Si--Sn
    alloy \cite{H-A}. Circles are taken from measurements performed on
    bulk single crystal Si(100), and squares are taken from
    measurements using Sn-implanted Si-on-sapphire (SOS) samples.
    Curves are given by the models for interfacial absolute stability:
    dashed - with local equilibrium diffusion and solute drag effect, 
    Eqs.~(\ref{eq:ms-VA}),
    (\ref{28}), and (\ref{26}); 
    solid - with the local nonequilibrium
    diffusion, Eqs.~(\ref{eq:ms-VA}), (\ref{25}), and (\ref{27}). 
    Dashed-dotted line, $V=V_D$, represents the limiting velocity 
    for the absolute interface stability.  }
  \label{fig:si_sn_va}
\end{figure}

%%%%%
\newpage
\begin{figure}[h]
\includegraphics[width=\textwidth]{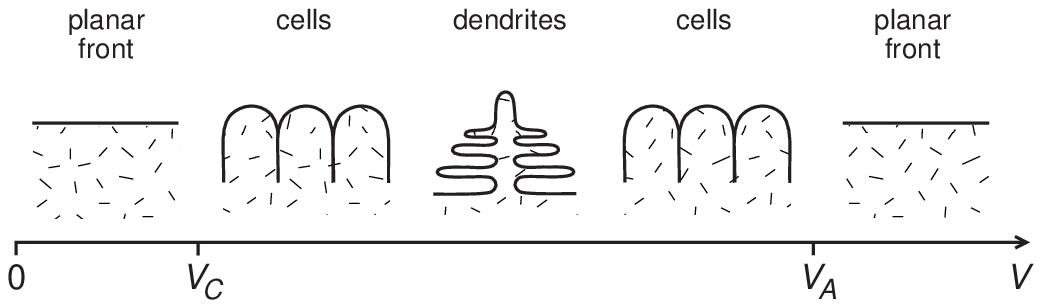}
\\
\center{Figure~\ref{fig:morph_spectr-1}}
\end{figure}

\newpage
\begin{figure}[h]
\includegraphics[width=1.1\textwidth]{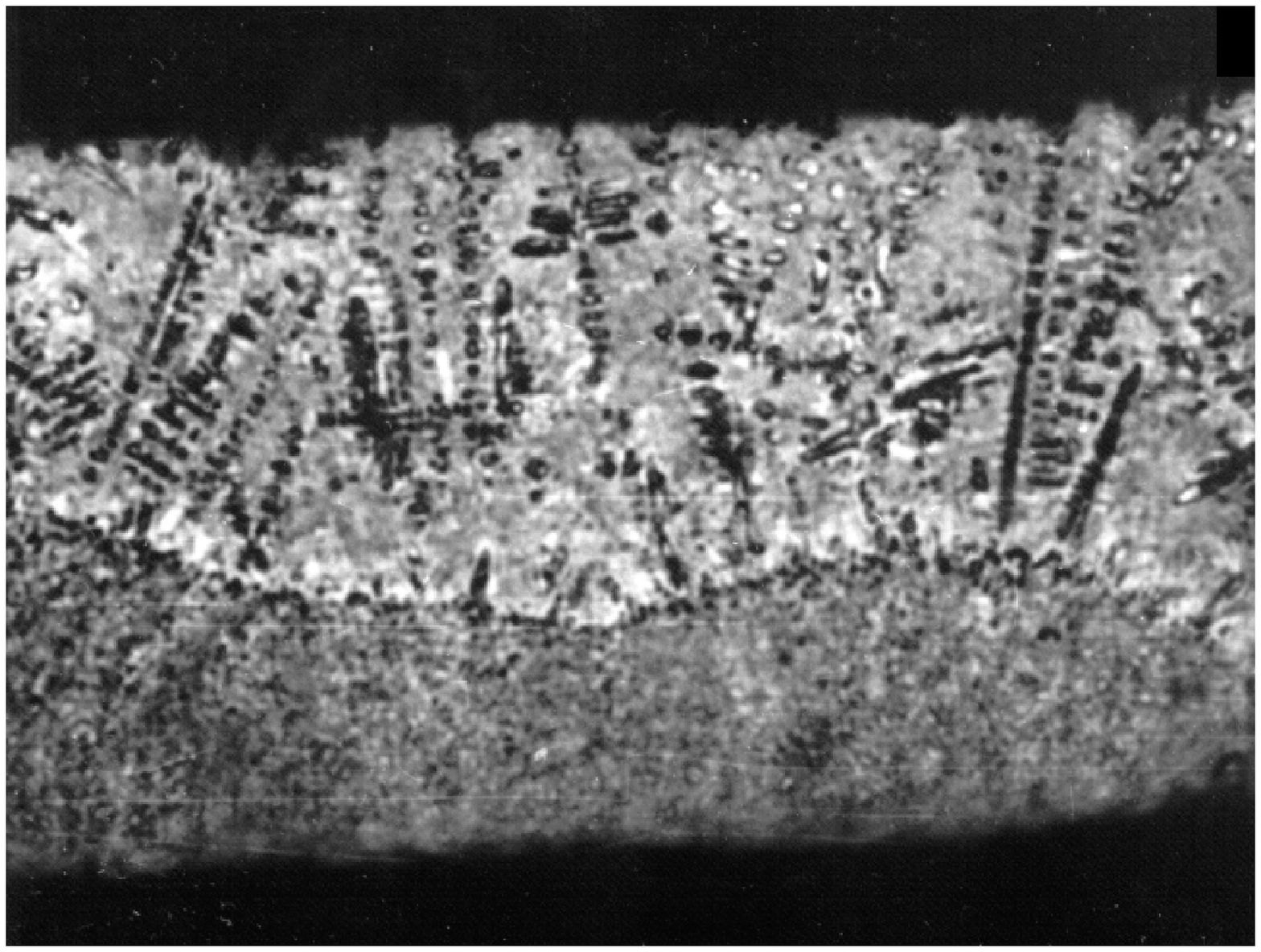}
\\
\center{Figure~\ref{fig:spinning}}
\end{figure}

\newpage
\begin{figure}[h]
  \includegraphics[width=\textwidth]{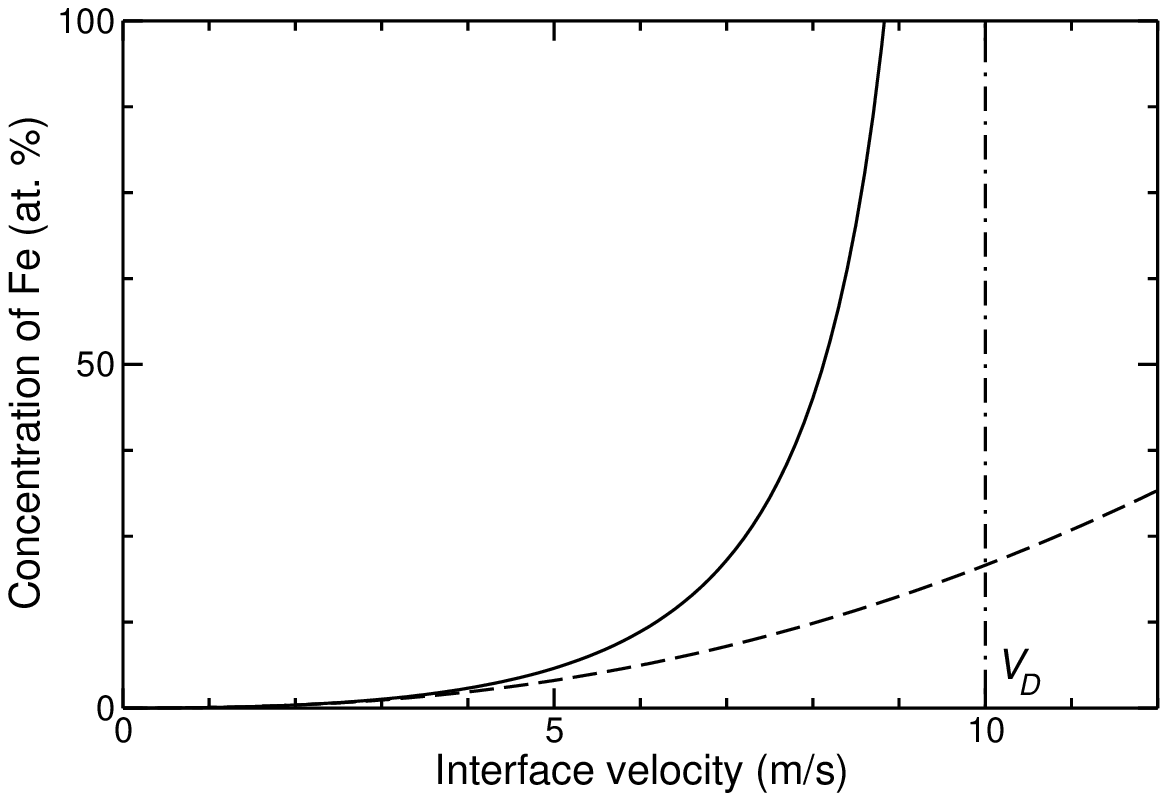}
\\
\center{Figure~\ref{fig:ms-VA-C0}} 
\end{figure}

\newpage
\begin{figure}[h]
  \includegraphics[width=1.1\textwidth]{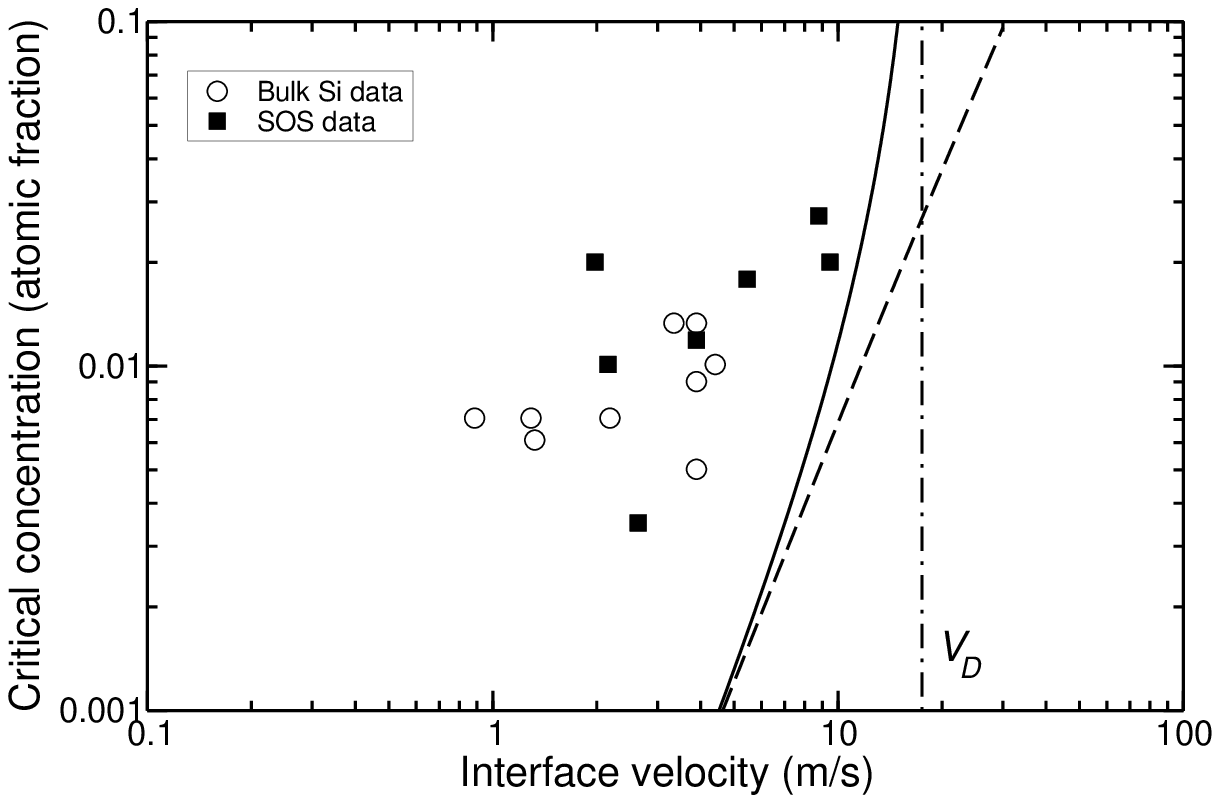}
\\\\\\\\
\center{Figure~\ref{fig:si_sn_va}}
\end{figure}

%\tableofcontents
\end{document}